\documentclass[twocolumn,showpacs,aps,floatfix,prd,nofootinbib]{revtex4}
\usepackage{graphicx}
\usepackage{dcolumn} 
\usepackage{amssymb} 
\usepackage{amsmath} 
\usepackage{colordvi}
\usepackage{color}   
\usepackage{hhline}  
\usepackage{psfrag} 

\RequirePackage{xspace}

\newcommand{\BABARPubYear}    {09}
\newcommand{\BABARPubNumber}  {006}
\newcommand{\SLACPubNumber} {13641}
 
\usepackage{relsize}
\def\babar{\mbox{\slshape B\kern-0.1em{\smaller A}\kern-0.1em
    B\kern-0.1em{\smaller A\kern-0.2em R}}}
     
\mathchardef\Upsilon="7107
\def\Y#1S{\ensuremath{\Upsilon{(#1S)}}\xspace}

\def\pep2{PEP-II}

\def\qbar  {\ensuremath{\overline q}\xspace}
\def\q     {\ensuremath{q}\xspace}

\long\def\inst#1{\par\nobreak\kern 4pt\nobreak
  {\it #1}\par\vskip 10pt plus 3pt minus 3pt}
  
\begin{document}

\begin{flushleft}
SLAC-PUB-\SLACPubNumber \\
\babar-PUB-\BABARPubYear/\BABARPubNumber \\
\end{flushleft}

\title{\large \bf
\boldmath
Measurement of the $\gamma\gamma^\ast \to \pi^0$
transition form factor}

\author{B.~Aubert}
\author{Y.~Karyotakis}
\author{J.~P.~Lees}
\author{V.~Poireau}
\author{E.~Prencipe}
\author{X.~Prudent}
\author{V.~Tisserand}
\affiliation{Laboratoire d'Annecy-le-Vieux de Physique des Particules (LAPP), Universit\'e de Savoie, CNRS/IN2P3,  F-74941 Annecy-Le-Vieux, France}
\author{J.~Garra~Tico}
\author{E.~Grauges}
\affiliation{Universitat de Barcelona, Facultat de Fisica, Departament ECM, E-08028 Barcelona, Spain }
\author{M.~Martinelli$^{ab}$}
\author{A.~Palano$^{ab}$ }
\author{M.~Pappagallo$^{ab}$ }
\affiliation{INFN Sezione di Bari$^{a}$; Dipartimento di Fisica, Universit\`a di Bari$^{b}$, I-70126 Bari, Italy }
% ...
\collaboration{The \babar\ Collaboration}
\noaffiliation

\begin{abstract}
We study the reaction $e^+e^-\to e^+e^-\pi^0$ 
and measure the $\gamma\gamma^\ast \to \pi^0$ transition form factor
in the momentum transfer range from 4 to 40 GeV$^2$. The analysis is
based on 442 fb$^{-1}$ of integrated luminosity collected at \pep2\ with 
the \babar\ detector at $e^+e^-$ center-of-mass energies near 10.6 GeV.
\end{abstract}

\pacs{14.40.Aq, 13.40.Gp, 12.38.Qk}

\maketitle

\setcounter{footnote}{0} 

\section{Introduction\label{intro}}
\begin{figure}
\includegraphics[width=.33\textwidth]{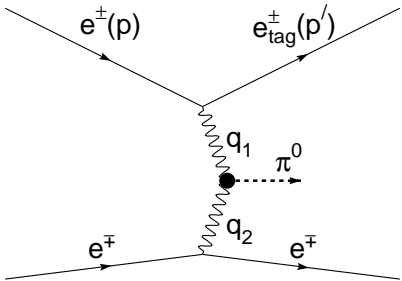}
\caption{The Feynman diagram for the $e^+e^-\to e^+e^-\pi^0$ two-photon 
production process.}
\label{fig1}
\end{figure}

In this paper we study the process 
\begin{equation}
e^+e^-\to e^+e^-\pi^0,
\end{equation}
where the final state $\pi^0$ is produced via the two-photon production
mechanism illustrated by Fig.~\ref{fig1}.
We measure the differential cross section for this process in the single
tag mode where one of the outgoing electrons\footnote{Unless otherwise
specified, we use the term ``electron'' for either an electron or a positron.}
 (tagged) is detected while the 
other electron (untagged) is scattered at a small angle. The $\pi^0$
is observed through its decay into two photons. The tagged electron emits
a highly off-shell photon with the momentum transfer 
$q^2_1 \equiv -Q^2 = (p-p^\prime)^2$, where
$p$ and $p^\prime$ are the four momenta of the initial and final electrons. 
The momentum transfer to the untagged electron is near zero. The differential
cross section for pseudoscalar meson production
$d\sigma(e^+e^-\to e^+e^-\pi^0)/dQ^2$ 
depends on only one form factor, $F(Q^2)$, which describes
the $\gamma\gamma^\ast \to \pi^0$ transition. To relate the differential 
cross section to the transition form factor we use the formulae 
for the $e^+e^-\to e^+e^-\pi^0$ cross section in Eqs.~(2.1) and (4.5) of 
Ref.~\cite{BKT}. 

At large momentum transfer, $Q^2$, perturbative QCD (pQCD) predicts that
the transition form factor can be represented as a convolution of a
calculable hard scattering amplitude for $\gamma\gamma^*\to q\bar{q}$
with a nonperturbative pion distribution amplitude, 
$\phi_\pi(x,Q^2)$~\cite{LB}.
The latter can be interpreted as the amplitude for the transition of the pion 
with momentum $P$ into two quarks with momenta $Px$ and $P(1-x)$.
In lowest order pQCD the transition form factor is obtained from:
\begin{eqnarray}
\lefteqn{Q^2F(Q^2)=} \nonumber\\
& & \frac{\sqrt{2}f_\pi}{3}\int_{0}^{1}\frac{dx}{x}\phi_\pi(x,Q^2)
+O(\alpha_s)+O(\frac{\Lambda^2_{\rm{QCD}}}{Q^2}),
\label{LO}
\end{eqnarray}
where $f_\pi=0.131$ GeV is the pion decay constant.
The pion distribution amplitude (DA) plays an important role in theoretical
descriptions of many hard-scattering QCD processes.
Since the evolution of $\phi_\pi(x,Q^2)$ with $Q^2$ 
is predicted by pQCD, experimental data on the transition form factor
can be used to determine its unknown dependence 
on $x$~\cite{th1,th2,th3,th4,th5,th6,th7,th8}.

The pion transition form factor was measured in the CELLO~\cite{CELLO}
and CLEO~\cite{CLEO} experiments
in the momentum transfer ranges 0.7--2.2 GeV$^2$ and 1.6--8.0 GeV$^2$,
respectively. In this paper we study the form factor
in the $Q^2$ range from 4 to 40 GeV$^2$. 

\section{ \boldmath The \babar\ detector and data samples}                 
\label{detector}                                                           
We analyze a data sample corresponding to an integrated
luminosity of about 442~fb$^{-1}$ recorded with                                                
the \babar\ detector~\cite{babar-nim} at the \pep2\                   
asymmetric-energy storage rings. At \pep2, 9-GeV electrons collide with     
3.1-GeV positrons to yield a center-of-mass energy of 10.58~GeV 
(the $\Upsilon$(4S) resonance). Additional data 
($\sim 43$~fb$^{-1}$) recorded at 10.54 GeV for the purpose of $\Upsilon$(4S)
background studies are included in the present analysis.

Charged-particle tracking is                                               
provided by a five-layer silicon vertex tracker (SVT) and                  
a 40-layer drift chamber (DCH), operating in a 1.5-T axial                 
magnetic field. The transverse momentum resolution                         
is 0.47\% at 1~GeV/$c$. Energies of photons and electrons                  
are measured with a CsI(Tl) electromagnetic calorimeter                    
(EMC) with a resolution of 3\% at 1~GeV. Charged-particle                  
identification is provided by specific ionization (${\rm d}E/{\rm d}x$)            
measurements in the SVT and DCH, and by an internally reflecting           
ring-imaging Cherenkov detector (DIRC). 

Signal and background $e^+e^-\to e^+e^-\pi^0\pi^0$ processes are simulated 
with the Monte Carlo (MC) event generator GGResRc.  It uses the formula
for the differential cross section from Ref.~\cite{BKT} for $\pi^0$
production and the BGSM formalism~\cite{Budnev} for the
two pion final state. Because the $Q^2$ distribution 
is peaked near zero, 
the MC events are generated with a restriction on the
momentum transfer to one of the electrons: $Q^2=-q_1^2 > 3$ GeV$^2$.
This restriction corresponds to the limit of detector acceptance for the
tagged electron. The second electron is required to have momentum
transfer $-q_2^2 < 0.6$ GeV$^2$. The experimental criterion providing this
restriction for data events is described in Sec.~\ref{evsel}.
The pseudoscalar form factor is fixed to $F(0)$ in MC simulation.

The GGResRc event generator includes next-to-leading-order radiative
corrections to the Born cross section calculated according to Ref.~\cite{RC}.
In particular, it generates extra soft photons emitted by 
the initial and final state electrons.
The formulae from Ref.~\cite{RC} were modified to account for
the hadron contribution to the vacuum polarization diagrams.
The maximum energy of the extra
photon emitted from the initial state is restricted by the 
requirement\footnote{Throughout this paper the asterisk
denotes quantities in the $e^+e^-$ c.m. frame.}
$E^\ast_\gamma < 0.05\sqrt{s}$, where $\sqrt{s}$ is the $e^+e^-$
center-of-mass (c.m.) energy.
The generated events are subjected to detailed
detector simulation based on GEANT4~\cite{GEANT4}, 
and are reconstructed with the
software chain used for the experimental data. Variations in the detector
and beam background conditions are taken into account.

Background events from $e^+e^-\to \q\qbar$,
where $\q$ represents a $u$, $d$, $s$ or $c$ quark, $e^+e^-\to \tau^+\tau^-$,
and $e^+e^-\to B\bar{B}$ 
are simulated with the JETSET~\cite{JETSET}, KK2F~\cite{KK2F}, and
EvtGen~\cite{EvtGen} event generators, respectively.

\section{Event selection\label{evsel}}
At the trigger level candidate events for the process under study are
selected by the {\tt VirtualCompton} filter.
This filter was originally designed to select so-called virtual
Compton scattering (VCS) events used for detector calibration.
This process corresponds to $e^+e^-\to e^+e^-\gamma$ with the 
kinematic requirement that 
one of the final state electrons goes along the collision axis, while the 
other electron and the photon are scattered at large angles.
The filter requires that a candidate event contain a track with 
$p^\ast/\sqrt{s} > 0.1$ and a cluster in the EMC with $E^\ast/\sqrt{s} > 0.1$
which is approximately opposite in azimuth ($|\delta{\phi^\ast}-\pi|<0.1$ rad)
to this track.
Cluster and track polar angle acolinearity in the c.m. frame is required to be
greater than 0.1 rad. Finally, the measured missing energy in the c.m. frame, 
which should correspond to the undetected electron, is compared to a prediction
based entirely on the directions of the detected particles, and the assumption 
that the missing momentum is directed along the collision axis:
$|E^\ast_{\rm{meas}}-E^\ast_{\rm{pred}}|/\sqrt{s} < 0.05$.
For a significant fraction of the $e^+e^-\to e^+e^-\pi^0$ events the trigger
cluster algorithm cannot separate the photons from $\pi^0$ decay, and hence 
identifies them as a single photon. Therefore the {\tt VirtualCompton} filter
has relatively large efficiency (about 50--80\% depending on the $\pi^0$ 
energy) for signal events.

In each event selected by the {\tt VirtualCompton} filter 
we search for an electron and a $\pi^0$ candidate.
A charged track identified as an electron must originate from the 
interaction point and be in the polar angle range $0.376 < \theta_e < 2.450$ 
rad. in the laboratory frame. The latter requirement is needed to provide high 
efficiency for the trigger track-finding algorithm and for good electron 
identification.
To recover electron energy loss due to bremsstrahlung,
both internal and in the detector material before the DCH,
we look for EMC showers close to the electron direction and combine their 
energies with the measured energy of the electron track. The resulting 
laboratory energy of the electron candidate must be greater than 2 GeV.
Two photon candidates with energies greater than 50 MeV are combined 
to form a $\pi^0$ candidate by requiring that their invariant mass be in the 
range 0.06--0.21 GeV/$c^2$ and that their laboratory energy sum be greater 
than 1.5 GeV.

The main background process, VCS, has a cross section several thousand times 
greater than that for the process under study. The VCS photon together
with a soft photon, for example from beam background, may give an invariant
mass value close to the $\pi^0$ mass. Such background events are effectively 
rejected by requirements on the photon helicity angle ($|\cos{\theta_h}|<0.8$)
and on the $\pi^0$ c.m. polar angle ($|\cos{\theta^\ast_\pi}|<0.8$). The photon
helicity angle $\theta_h$ is defined as the angle
between the decay photon momentum in the $\pi^0$ rest frame and the $\pi^0$ 
direction in the laboratory frame.

The next step is to remove improperly reconstructed QED events. We remove 
events which involve noisy EMC channels, events with extra tracks close to 
the $\pi^0$ candidate direction, and events with 
$|\Delta\theta_{\gamma\gamma}|<0.025$ rad, where $\Delta\theta_{\gamma\gamma}$ 
is the difference between the laboratory polar angles of
the photons from the $\pi^0$ decay. The latter condition removes
VCS events where the photon converted to an $e^+e^-$ pair within the DCH volume.
It also removes about 20\% of the signal events, but significantly improves 
(by a factor of about 15) the signal-to-background ratio.

\begin{figure}
\includegraphics[width=.48\textwidth]{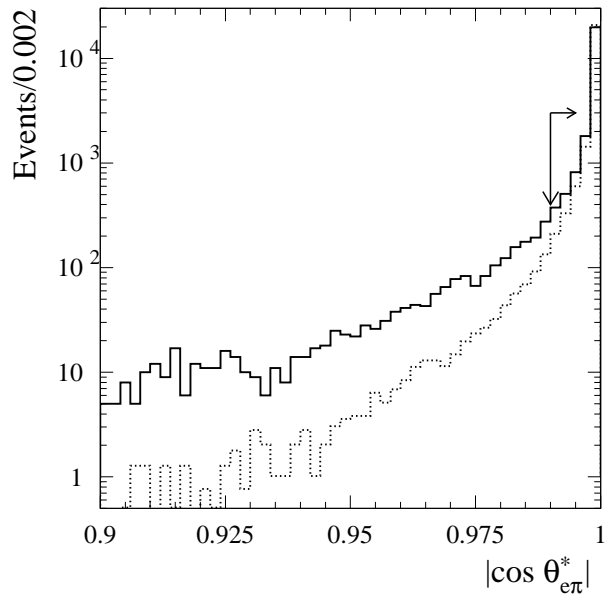}
\caption{The distribution of the cosine of the
polar angle of the $e\pi^0$ system momentum in the c.m. frame
for data (solid histogram) and simulated signal
(dotted histogram). Events for which  $|\cos{\theta^\ast_{e\pi}}|>0.99$
(indicated by the arrow) are retained.
\label{fig2}}
\end{figure}
Two additional event kinematics requirements
provide further background suppression and improved 
data to MC-simulation correspondence.
Figure~\ref{fig2} shows the data and MC simulation distributions of the
cosine of the polar angle of the momentum vector of the $e\pi^0$ system
in the c.m. frame.
We require $|\cos{\theta^\ast_{e\pi}}|>0.99$. This effectively limits 
the value of the momentum transfer to the untagged electron ($q^2_2$) and
guarantees compliance with the condition $-q^2_2<0.6$ GeV$^2$ used in MC 
simulation.

\begin{figure}
\includegraphics[width=.48\textwidth]{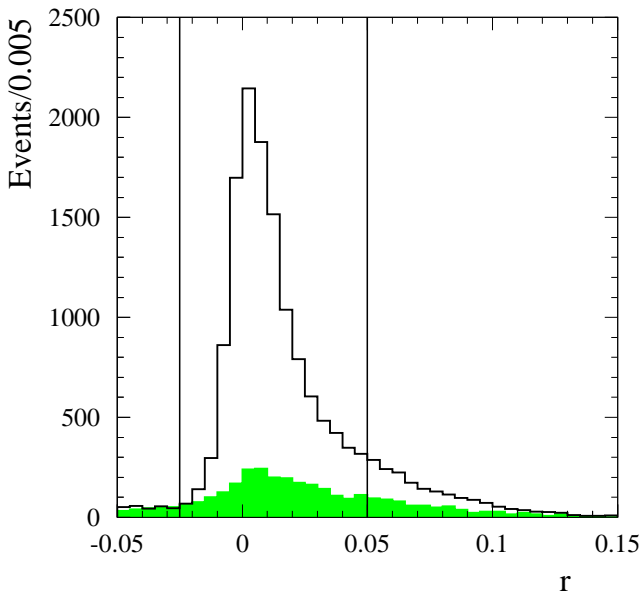}
\caption{The $r$ distribution for data events.
The shaded histogram shows 
the background contribution estimated from
the fit to the two-photon mass distribution (Sec.~\ref{fitting}). 
The vertical lines indicate the region used to select candidate events
($-0.025<r<0.050$).
\label{fig3}}
\end{figure}
The emission of extra photons by the electrons involved leads to a difference 
between the measured and actual values of $Q^2$. In the case of initial state 
radiation (ISR) $Q^2_{\rm{meas}}=Q^2_{\rm{true}}(1+r_\gamma)$, where 
$r_\gamma=2E^\ast_\gamma/\sqrt{s}$.
To restrict the energy of the ISR photon we use the parameter
\begin{equation}
r=\frac{{\sqrt{s}}-E^\ast_{e\pi}-p^\ast_{e\pi}}{\sqrt{s}},
\label{eqr} 
\end{equation}
where $E^\ast_{e\pi}$ and $p^\ast_{e\pi}$ are the c.m. energy and the
magnitude of the momentum 
of the detected $e\pi^0$ system. In the ISR case this parameter   
coincides with $r_\gamma$ defined above. 
The condition $r<0.075$ ensures compliance with the restriction
$r_{\gamma}<0.1$ used in MC simulation. The $r$ distribution for data
is shown in Fig.~\ref{fig3}, where the shaded histogram shows the background 
estimated from the fit to the two-photon mass distribution (Sec.~\ref{fitting}).
We select events with $-0.025<r<0.050$ for further analysis.

The background from $e^+e^-$ annihilation into hadrons is strongly
suppressed by the requirements of electron identification,    
on $\cos{\theta^\ast_{e\pi}}$, and on $r$. An additional
two-fold suppression of this background is provided by the condition that 
the $z$-component of the c.m. momentum of the $e\pi^0$ system is negative 
(positive) for events with a tagged positron (electron).

\begin{figure}
\includegraphics[width=.48\textwidth]{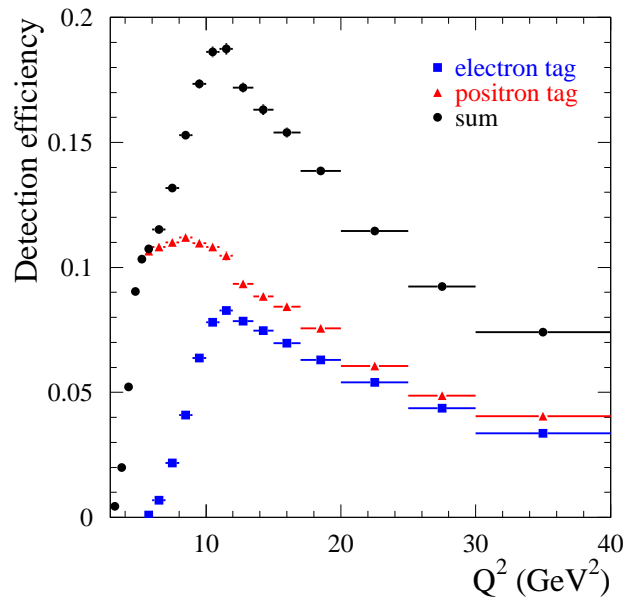}
\caption{The detection efficiency as a function of the 
momentum transfer squared for events with a tagged electron (squares),
a tagged positron (triangles), and their sum (circles).
\label{fig4}}
\end{figure}
The $Q^2$ dependence of the detection efficiency obtained from 
MC simulation is shown in Fig.~\ref{fig4}.
The detector acceptance limits the detection efficiency at small $Q^2$.
To avoid possible systematics due to data-simulation differences near
detector edges, we measure the cross section and form factor in the region
$Q^2 > 4$ GeV$^2$. The asymmetry of the $e^+e^-$ collisions at PEP-II
leads to different efficiencies for events with electron and positron
tags. The $Q^2$ range from 4 to 7 GeV$^2$ is measured only with
the positron tag.
The decrease of the detection efficiency in the region $Q^2 > 10$ GeV$^2$ is
explained by the decrease of the $\pi^0$ reconstruction efficiency due
to growth of the average $\pi^0$ energy with $Q^2$.

\section{Fitting the two-photon mass spectrum}\label{fitting}
\begin{figure}
\includegraphics[width=.48\textwidth]{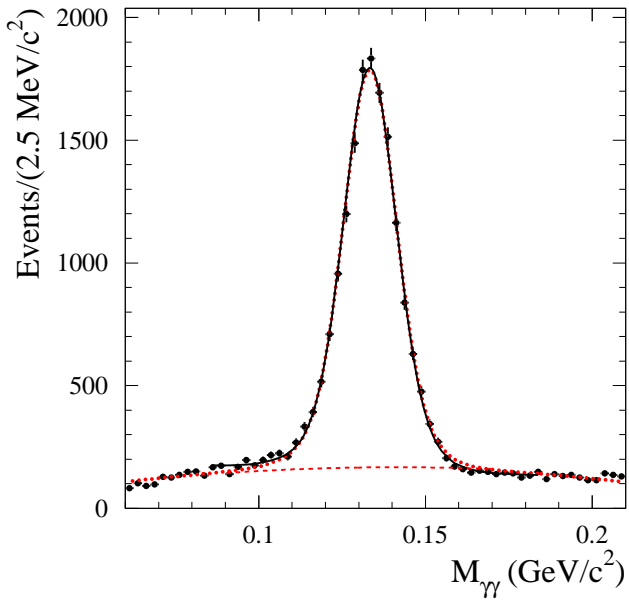}
\caption{The two-photon invariant mass spectrum for data events
with $4<Q^2<40$ GeV$^2$. The solid (dotted) 
curve corresponds to the fit with a linear (quadratic) background shape.
The dashed curve represents the fitted quadratic background.
\label{fig5}}
\end{figure}

The two-photon mass spectrum for selected data events with 
$4 < Q^2 < 40$ GeV$^2$ is shown in Fig.~\ref{fig5};
for $Q^2 >40 $ GeV$^2$ we do not see evidence of a $\pi^0$ signal 
over background. To determine the number of events
containing a $\pi^0$ we perform a binned likelihood fit to the spectrum
with a sum of signal and background distributions. 
We describe
the signal line shape by a sum of two $F_{\rm{B1}}$ functions with 
the same position of their maxima~\cite{bukin}. The function
$F_{B1}$ is the convolution of a Gaussian and an exponential
distribution:
\begin{multline}
{F_{\rm{B1}}(x;x_g,\sigma_g,\lambda)=
\frac{1}{2|\lambda|}\exp \left (
-\frac{x-x_g}{\lambda}+\frac{\sigma_g^2}{2\lambda^2} 
\right )\times}\\
\left[1-\mathrm{erf} \left(
\frac{\sigma_g^2-(x-x_g)\lambda}{\sqrt{2}\sigma_g|\lambda|} \right)
\right ].
\end{multline}

The parameters of the $\pi^0$ resolution function are fixed
from the fit of the mass spectrum obtained for simulated signal events
weighted to yield the $Q^2$ dependence observed in data.
The background distribution is described either by a linear function in the 
mass range 0.085--0.185 GeV/$c^2$ or a second order polynomial in the mass 
range 0.06--0.21 GeV/$c^2$.
The data mass spectrum is fitted with 5 (6 for second order polynomial) free
parameters: the number of signal events, the peak position, 
the sigma of one (narrow) of the $F_{\rm{B1}}$ functions ($\sigma_1$),
and 2 (3) parameters for the background.
The results of the fits are shown in Fig.~\ref{fig5}.

The total number of signal events is about 14000. The difference in signal
yield between the two background hypotheses is 170 events, while the 
statistical error on the signal yield is 140 events. The difference between 
the peak positions in data and MC simulation is consistent with zero. The 
value of $\sigma_1$ is 7.5 MeV/$c^2$ in data and 7.7 MeV/$c^2$ in simulation,
which corresponds to a difference of about two standard deviations.

\begin{figure*}
\includegraphics[width=.32\textwidth]{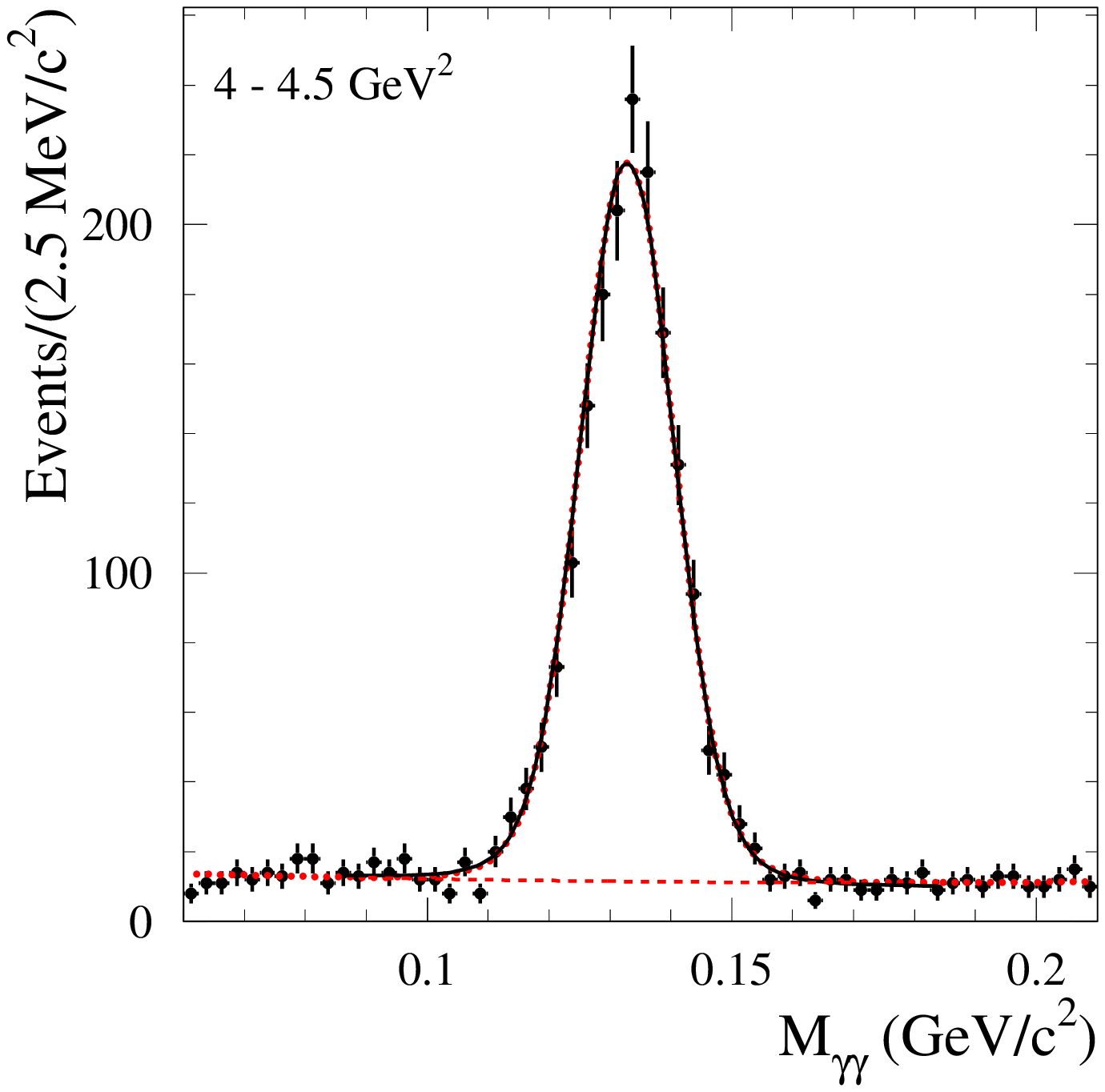}
\includegraphics[width=.32\textwidth]{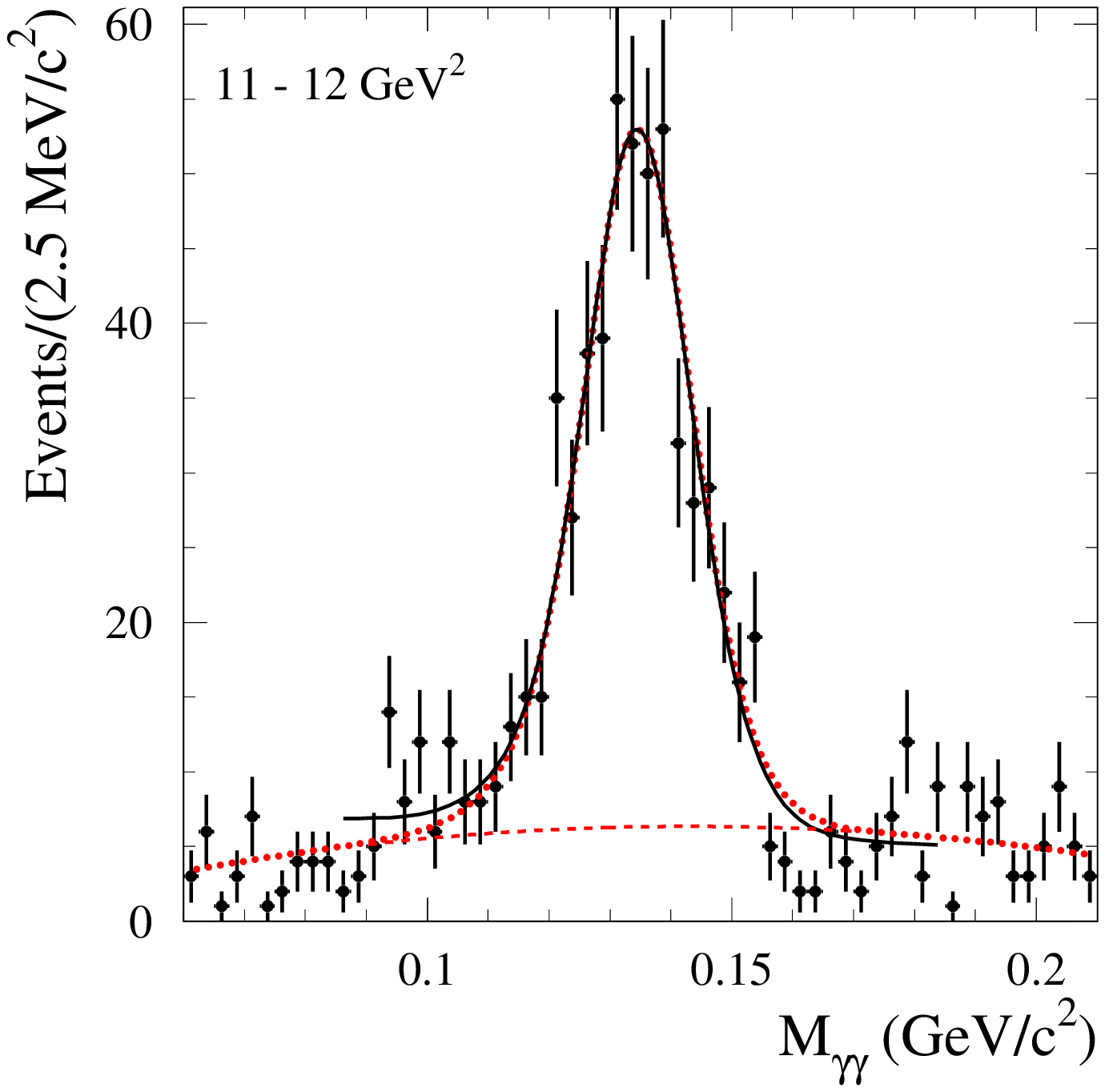}
\hfill
\includegraphics[width=.32\textwidth]{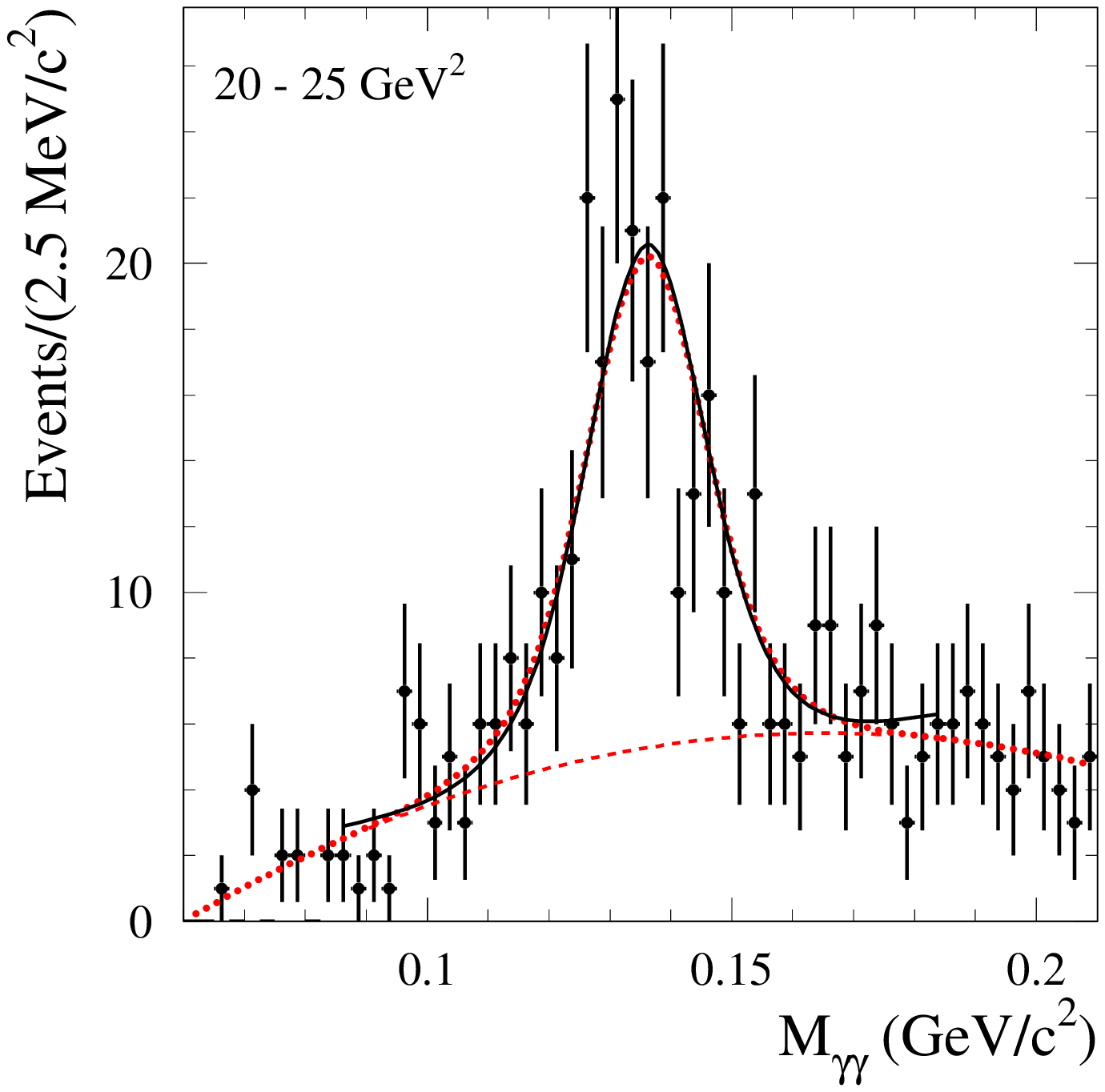}
\caption{The two-photon invariant mass spectra for data events
from three representative $Q^2$ intervals.
The solid (dotted) 
curve corresponds to the fit with a linear (quadratic) background shape.
The dashed curve represents the fitted quadratic background.
\label{fig6}}
\end{figure*}
A similar fitting procedure is applied in each of the seventeen $Q^2$
intervals indicated in Table~\ref{tab10}. The parameters of the $\pi^0$
resolution function are taken from the fit of the mass spectrum for
simulated events in the corresponding $Q^2$ interval. For the fits to the data,
the value of the parameter $\sigma_1$ is modified to take into
account the observed data-simulation difference in resolution:
$\sigma_1 \rightarrow \sqrt{\sigma_1^2-(1.9\mbox{ MeV})^2}$.
The free parameters in the data fits are the number of signal events and
two or three parameters, depending upon the description of the background 
shape.
The numbers of signal events obtained from the fits using a linear background
are listed in Table~\ref{tab10}. The difference between the fits for the two 
background hypotheses is used as an estimate of the systematic uncertainty 
associated with the unknown background shape. The two-photon mass spectra and 
fitted curves for three representative $Q^2$ intervals are shown
in Fig.~\ref{fig6}.
\begin{table*}
\caption{For each $Q^2$ interval, the number of events with $\pi^0$ obtained 
from the
fit ($N_\pi$), number of $e^+e^-\to e^+e^-\pi^0\pi^0$ background events 
($N_{\rm{bkg}}$), total efficiency correction 
($\delta_{\rm{total}}$), number of signal events corrected for data/MC 
difference and resolution effects ($N_{\rm{cor}}$), 
and detection efficiency obtained from 
simulation ($\varepsilon$). The quoted errors on $N_\pi$ and $N_{\rm{cor}}$ are 
statistical and systematic. For $N_{\rm{cor}}$ we quote only $Q^2$-dependent
systematic errors. The $Q^2$-independent systematic error is 2.5\%.
\label{tab10}\\}
\begin{ruledtabular}
\newcolumntype{+}{D{+}{\,\pm\,}{-1}}
\begin{tabular}{d++++d}
\multicolumn{1}{c}{$Q^2$ interval (GeV$^2$)} &
\multicolumn{1}{c}{$N_\pi$}      & 
\multicolumn{1}{c}{$N_{\rm{bkg}}$}  &
\multicolumn{1}{c}{$\delta_{\rm{total}} (\%)$} &
\multicolumn{1}{c}{$N_{\rm{cor}}$}    &
\multicolumn{1}{c}{$\varepsilon$ (\%)}  \\
\hline
 4.0-4.5  & 1645+45\pm4 &176+41 &-4.9+1.2 &1503+52\pm52 &  5.2 \\ 
 4.5-5.0  & 1920+49\pm11 &254+54 &-5.5+1.1 &1740+58\pm70 &  9.0 \\
 5.0-5.5  & 1646+46\pm5 &206+34 &-5.0+1.1 &1551+56\pm46 & 10.3 \\
 5.5-6.0  & 1252+41\pm5 &175+30 &-5.5+1.0 &1139+50\pm40 & 10.7 \\
 6.0-7.0  & 1891+50\pm2 &271+36 &-7.0+1.1 &1760+59\pm47 & 11.5 \\
 7.0-8.0  & 1229+41\pm19 &150+29 &-7.5+1.0 &1160+50\pm44 & 13.2 \\
 8.0-9.0  &  985+38\pm27 &125+24 &-7.3+0.9 & 915+46\pm46 & 15.3 \\
 9.0-10.0 &  829+34\pm8 & 59+14 &-7.7+1.0 & 849+43\pm23 & 17.3 \\
10.0-11.0 &  625+30\pm18 & 47+13 &-8.3+1.1 & 634+40\pm30 & 18.6 \\
11.0-12.0 &  448+26\pm3 & 27+11 &-8.4+1.0 & 484+35\pm16 & 18.7 \\
12.0-13.5 &  405+26\pm22 & 51+12 &-8.1+0.9 & 381+33\pm32 & 17.2 \\
13.5-15.0 &  289+22\pm14 & 13+ 6 &-7.3+1.0 & 304+28\pm20 & 16.3 \\
15.0-17.0 &  260+22\pm5 & 14+ 6 &-6.7+1.0 & 270+27\pm11 & 15.4 \\
17.0-20.0 &  235+21\pm2 & 20+ 6 &-6.6+1.1 & 234+25\pm10 & 13.9 \\
20.0-25.0 &  171+19\pm11 &  5+ 4 &-6.6+1.3 & 185+22\pm14 & 11.4 \\
25.0-30.0 &   36+12\pm2 &  1+ 1 &-6.9+1.5 &  36+14\pm3 &  9.2 \\
30.0-40.0 &   49+12\pm2 &  2+ 6 &-6.3+1.8 &  53+13\pm8 &  7.3 \\
\end{tabular}
\end{ruledtabular}
\end{table*}

\section{Peaking background estimation and subtraction}\label{background}
\begin{figure}
\includegraphics[width=.48\textwidth]{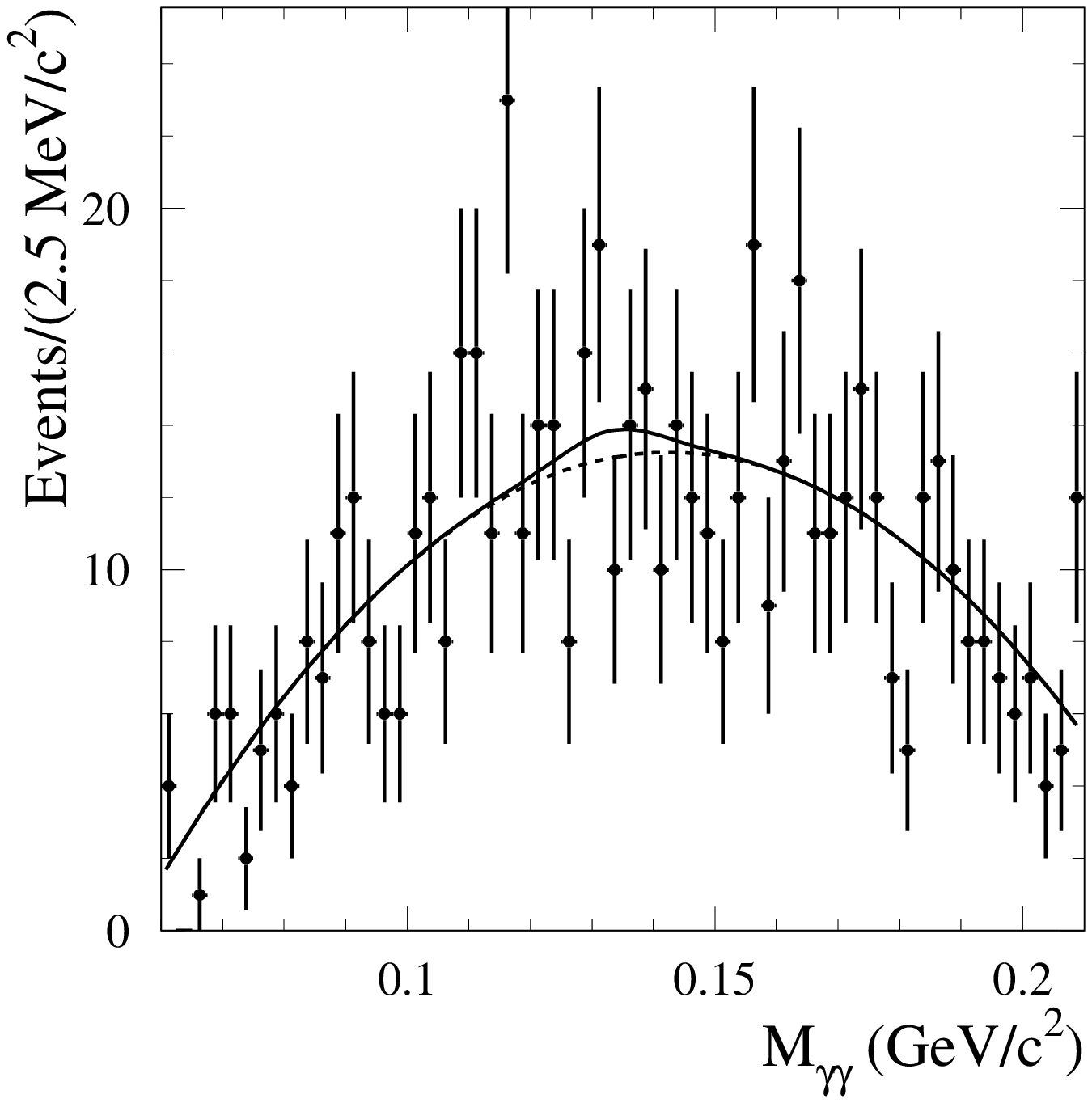}
\caption{The two-photon mass spectrum for
data with the wrong sign of the $e\pi^0$ c.m. momentum component along the
collision axis. The solid curve is the result of the fit described in the text.
The dashed curve represents the fitted non-peaking background.
\label{fig7}}
\end{figure}
Background containing true $\pi^0$'s might arise from processes such as 
beam-gas interaction, $e^+e^-$ annihilation, and two-photon processes 
yielding higher multiplicity final states. 

For beam-gas interactions, the total energy of the detected electron and 
$\pi^0$ should be less than the beam energy. In the energy spectrum
of the $e\pi^0$-system we do not see events with energy less than 
the beam energy. Therefore we conclude that beam-gas background does not 
survive the selection criteria.

For events due to the signal process with tagged positron (electron),
the momentum of the detected $e\pi^0$ system in the $e^+e^-$ c.m. frame
has negative (positive) $z$-component, while events resulting from $e^+e^-$ 
annihilation should be produced symmetrically.
Events with the wrong sign of the $e\pi^0$ momentum $z$-component
can therefore be used to estimate the background contribution from 
$e^+e^-$ annihilation.
The two-photon mass spectrum for such background events is shown in 
Fig.~\ref{fig7}. 
The total number of wrong-sign events is about 3\% of the selected signal
event candidates. The spectrum is fitted using a sum of signal and background
distributions as described in Sec.~\ref{fitting}.
The fit yields $6\pm16$ $\pi^0$ events.
Assuming that the numbers of background events from
$e^+e^-$ annihilation in the wrong and right-sign data samples are 
approximately the same, we conclude that this background does not exceed
0.2\% of signal events, and so is negligible.
Nevertheless we have analyzed simulated events for the processes 
$e^+e^-\to B\bar{B}$, $e^+e^-\to q\bar{q}$, and $e^+e^-\to \tau^+\tau^-$.
The number of simulated events for each reaction is close to
the number of such events produced in the experiment. 
Seven events with right-sign $z$-component of the $e\pi^0$ momentum (4 from
$\tau^+\tau^-$ and 3 from $q\bar{q}$) satisfy the analysis selection criteria,
while the number of accepted wrong-sign events is four.
This supports the conclusion that the $e^+e^-$ annihilation background is
negligible.
\begin{figure}
\includegraphics[width=.48\textwidth]{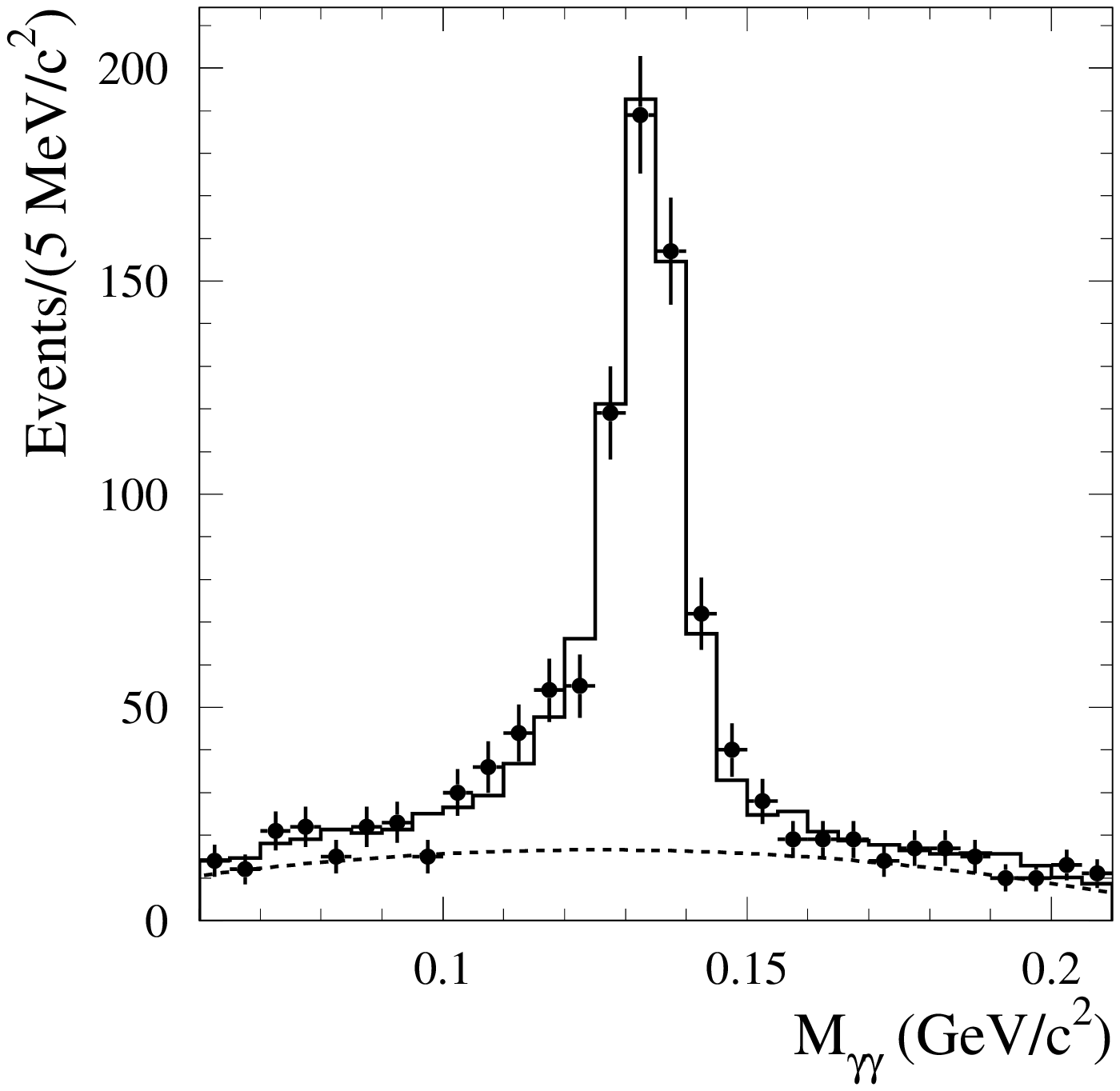}
\caption{The invariant mass spectrum of the extra $\pi^0$ candidate in
$2\pi^0$
data events (points with error bars). The histogram is the result
of the fit using a sum of signal and background distributions. The dashed
curve represents the fitted background distribution.
\label{fig8}}
\end{figure}
\begin{figure}
\includegraphics[width=.48\textwidth]{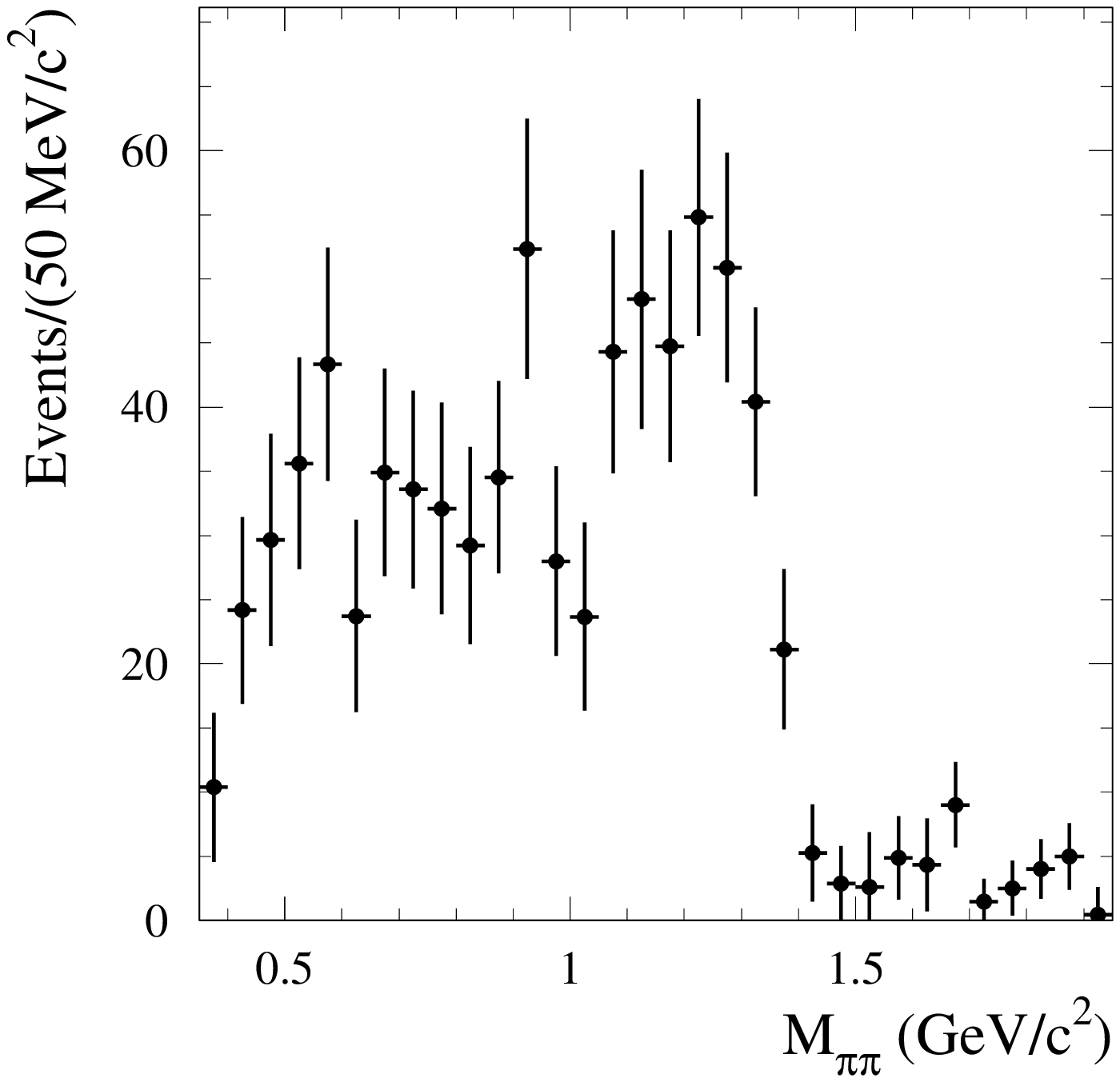}
\caption{The $2\pi^0$ invariant mass spectrum for data.
\label{fig9}}
\end{figure}

The major source of peaking background, of order 10\%, 
is two-photon production of two $\pi^0$'s. 
This background is clearly seen as a $\pi^0$ peak
in the two-photon invariant mass spectrum for data events with
two extra photons. The following procedure is used to estimate the
$2\pi^0$ background. We select a clean sample of $2\pi^0$ events with the
special selection criteria (described below) and measure the $Q^2$ distribution 
for these events ($N_{2\pi,i}$). Then we tune the MC simulation
of the $e^+e^-\to e^+e^-\pi^0\pi^0$ reaction to reproduce the $2\pi^0$ mass
and $\pi^0$ angular distributions observed in data. 
Using the MC simulation we calculate the ratio ($\kappa_i$) of the numbers of
$2\pi^0$ events selected with the standard and special criteria 
and estimate the number of $2\pi^0$ events for each $Q^2$ interval
that satisfy the standard selection criteria  as $\kappa_i N_{2\pi,i}$.

To select $2\pi^0$ events we remove the criteria on
$r$ and $\cos{\theta^\ast_{e\pi}}$ and search for events with an extra
$\pi^0$. Combinatorial background due to soft false photons is reduced
by requiring that, in the laboratory, the energy of the extra $\pi^0$ 
be greater than 0.2 GeV,
and that the energies of the decay photons be greater than 50 MeV.
The mass of the first $\pi^0$ must be in the range 0.10--0.17 GeV/$c^2$.
We calculate the parameters $\cos{\theta^\ast}$ and $r$ for
the found $e\pi^0\pi^0$ system, and require 
$|\cos{\theta^\ast_{e\pi\pi}}| > 0.99$ and $-0.025<r<0.05$.
The two-photon invariant mass spectrum for the extra $\pi^0$ candidates is 
shown in Fig.~\ref{fig8}. The mass spectrum is fitted using a sum of signal and
background distributions. The signal distribution is obtained from
MC simulation for the $e^+e^- \to e^+e^-\pi^0\pi^0$ process. The background is
described by a second order polynomial. This fitting procedure is
performed for all $Q^2$ intervals. To estimate the systematic uncertainty due
to the unknown background shape we make two fits, one with a linear, and one with
a quadratic background. The difference in $\pi^0$ signal size between the fits
is taken as a measure of systematic uncertainty.

\begin{figure}
\includegraphics[width=.48\textwidth]{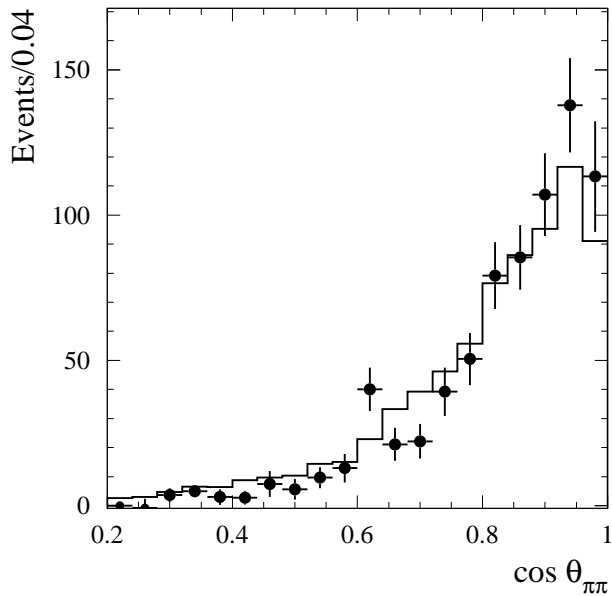}
\caption{The distribution of $\cos{\theta_{\pi\pi}}$ for data events
(points with error bars) and simulated $e^+e^-\to e^+e^-\pi^0\pi^0$ events
with isotropic $\pi^0$ angular distribution (histogram).
\label{fig10}}
\end{figure}

The $2\pi^0$ mass spectrum for selected $2\pi^0$ events after background
subtraction is shown in Fig.~\ref{fig9}. The observed spectrum 
differs strongly from the spectrum measured in the no-tag 
mode~\cite{Belle2pi0}, where the dominant mechanism of $2\pi^0$
production is $\gamma\gamma\to f_2(1270) \to \pi^0\pi^0$. In the no-tag
mode the $f_2(1270)$ meson is produced predominantly in the helicity $2$ state
with angular distribution $\sim\sin^4\theta_{\pi\pi}$,
where the $\theta_{\pi\pi}$ is the angle between the $\pi^0$ direction
and the assumed $\gamma\gamma$ collision axis in the di\-pion rest frame.
The $\cos{\theta_{\pi\pi}}$ distribution for selected $2\pi^0$ events
after background subtraction is shown in Fig.~\ref{fig10}. It is seen
that our criteria select events with $\theta_{\pi\pi}$ near zero
and strongly suppress $f_2(1270)$ production in the helicity-$2$ state.
The spectrum in Fig.~\ref{fig9} contains three components:
tensor $f_2(1270)$, scalar $f_0(980)$, and a broad bump below 0.8 GeV/$c^2$.
We reweight the simulated events to reproduce the mass
spectrum observed in data. Since the mass spectrum may change with
$Q^2$, the reweighting is performed for two $Q^2$ intervals 
($4 < Q^2 < 10$ GeV$^2$ and $10 < Q^2 < 40$ GeV$^2$) separately.   

The simulated events are generated with isotropic $\pi^0$ angular distribution
in the $2\pi^0$ rest frame.
Comparison of the simulated $\cos{\theta_{\pi\pi}}$ distribution
with the data distribution is shown in Fig.~\ref{fig10}.
We reweight the $f_2(1270)$ subsample of simulated events 
so that the total MC simulated distribution of Fig.~\ref{fig10}
matches the data.
Using reweighted simulated events we calculate 
the $Q^2$ dependence of the scale factor $\kappa_i$ which varies from 2.4
at $Q^2\sim5$ GeV$^2$ to about 1 at $Q^2>15$ GeV$^2$.
The numbers of $2\pi^0$ background events which satisfy our standard 
selection criteria are listed in Table~\ref{tab10}. The fraction of $2\pi^0$ 
background events in the $e\pi^0$ data sample changes from about 13\% 
for $Q^2 <10$ GeV$^2$ to 6--7\% for $Q^2 > 10$ GeV$^2$.

A similar technique is used to search for background from the processes
$e^+e^-\to e^+e^-\pi^0\eta,\, \eta\to\gamma\gamma$ and 
$e^+e^-\to e^+e^-\omega,\, \omega\to\pi^0\gamma$.
We do not see a clear $\eta$ signal in the two-photon mass spectrum, nor
do we see an $\omega$ signal in the $\pi^0\gamma$ mass spectrum; 
we estimate that these backgrounds do not exceed 5\% of the $2\pi^0$ 
background and thus are negligible.

\section{Efficiency correction\label{effcor}}
The values of the $\gamma^\ast\gamma\to\pi^0$ transition form factor in
bins of $Q^2$ are
determined from the ratio of the $Q^2$ distributions from data and MC simulation.
The data distribution must be corrected to account for data-simulation
difference in detector response:
\begin{equation}
N_i^{corr}=\frac{N_i}{\prod_{j=1}^{4}(1+\delta_j^i)},
\label{eqcor}
\end{equation}
where $i$ denotes the interval of $Q^2$ under consideration, and
the $\delta_j^i$'s are the corrections for the effects discussed 
in Secs.~\ref{effcor1}-\ref{effcor4}.
\subsection{$\pi^0$ reconstruction efficiency\label{effcor1}}
A possible source of data-simulation difference is $\pi^0$ loss
due to the merging of electromagnetic showers produced by the two photons 
from the $\pi^0$ decay, the loss of at least one of the decay photons,
or the rejection of the $\pi^0$ because of the selection 
criteria. The $\pi^0$ efficiency is studied by using events produced in the ISR
process $e^+e^-\to \omega\gamma$, where $\omega\to\pi^+\pi^-\pi^0$~\cite{omega1}. 
These events can be selected and reconstructed
using the measured parameters for only the two charged tracks and the ISR 
photon. Taking the ratio of the number of events with found $\pi^0$ to the 
total number of selected $e^+e^-\to \omega\gamma$ events we measure 
the  $\pi^0$ reconstruction efficiency. 
The events with reconstructed $\pi^0$ are selected with our
standard criteria for the photons and the $\pi^0$, as described in
Sec.~\ref{evsel}.

\begin{figure}
\includegraphics[width=.48\textwidth]{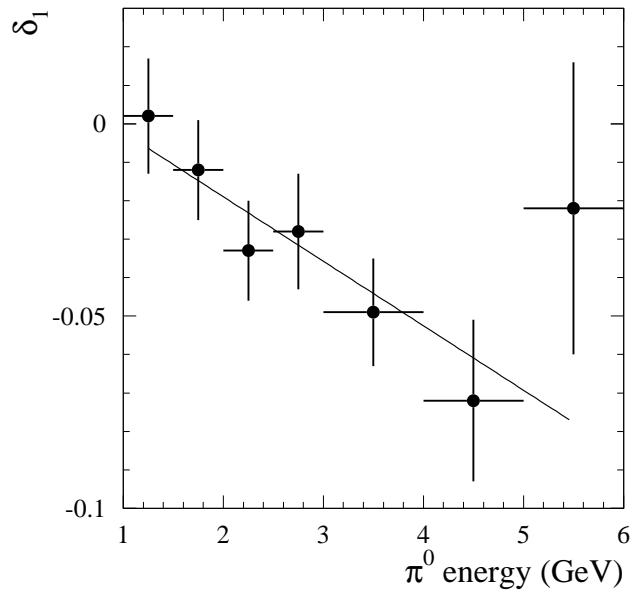}
\caption{The correction to the MC-estimated $\pi^0$ efficiency as a function 
of $\pi^0$ energy in the laboratory frame.
\label{fig11}}
\end{figure}
The ratio of the reconstruction efficiencies 
obtained in data and in simulation provides a $\pi^0$ efficiency correction. 
This correction, $\delta_1=\epsilon_{\rm{data}}/\epsilon_{\rm{MC}}-1$, is shown 
as a function of $\pi^0$ laboratory energy in  Fig.~\ref{fig11}.
The energy dependence is well described by a linear function.
We estimate that the systematic uncertainty associated with this correction 
does not exceed 1\%.

\begin{figure}
\includegraphics[width=.48\textwidth]{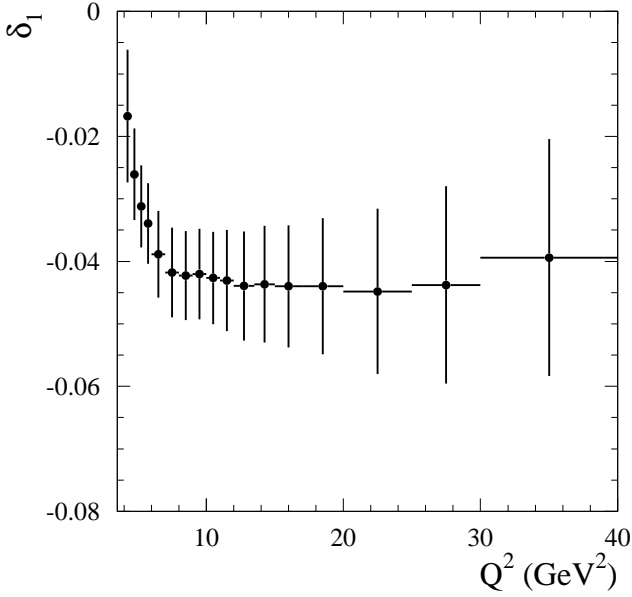}
\caption{The $Q^2$ dependence of the correction to the MC-estimated 
efficiency for $\pi^0$.
\label{fig12}}
\end{figure}
To obtain the correction to the MC-estimated $\pi^0$ efficiency 
as a function of $Q^2$ 
we convolve the correction energy dependence of Fig.~\ref{fig11} with the 
$\pi^0$ energy spectrum in each $Q^2$ interval.
The $Q^2$ dependence obtained is shown in Fig.~\ref{fig12}. Only statistical
errors are shown; the systematic error is estimated to be 1\%.

\begin{figure}
\includegraphics[width=.48\textwidth]{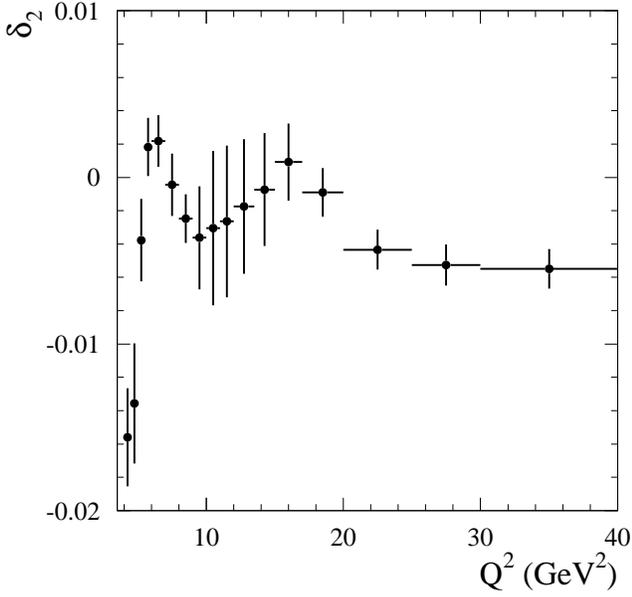}
\caption{The correction to the MC-estimated EID efficiency as a function
of $Q^2$.
\label{fig13}}
\end{figure}
\subsection{Electron identification efficiency\label{effcor2}}
The average electron identification (EID) inefficiency 
in the signal MC simulation is about 1\%.
To estimate the data-simulation difference in EID we
use VCS events which can be selected with negligible background without
any EID requirements.  The EID efficiency is
determined as the ratio of the number of events with an identified electron
to the total number of VCS events. The ratio of
the efficiencies obtained from data and simulation gives the efficiency
correction. We determine the correction as a function of
the electron energy and polar angle and convolve this function
with the electron energy and angular distributions for the
process under study.  The resulting $Q^2$ dependence of the efficiency 
correction is shown in Fig.~\ref{fig13}. 

\subsection{Trigger efficiency\label{effcor3}}
With the available statistics and the trigger configuration used, we
cannot determine the trigger efficiency for the process under study by using
the data.  However the trigger efficiency can be measured for the 
VCS process, which has a much larger cross section. 
The VCS events allow the determination of the part of the trigger inefficiency 
related to the trigger track-finding algorithm. 
The remaining trigger inefficiency, which is related to the ability of the 
trigger cluster algorithm to separate nearby photons from $\pi^0$ decay,
depends strongly on $\pi^0$ energy. Therefore the data-simulation
difference can be estimated from comparison of the $\pi^0$ energy spectra in 
data and simulation.
\begin{figure}
\includegraphics[width=.48\textwidth]{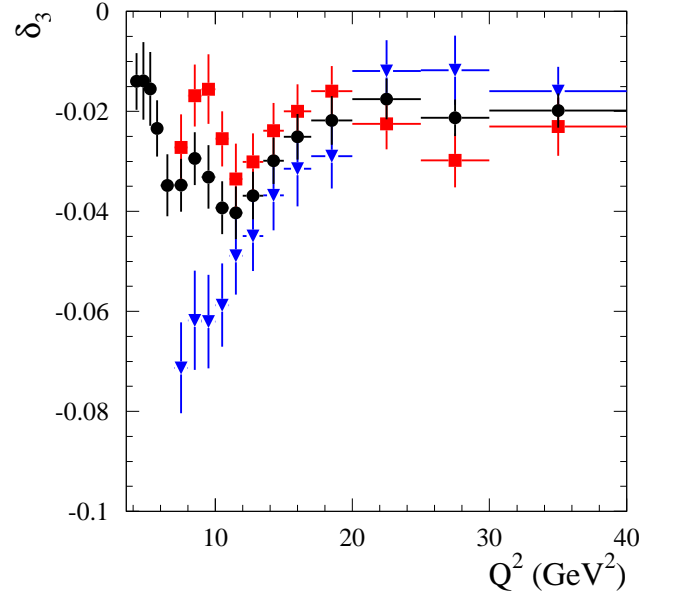}
\caption{
The $Q^2$ dependence of the correction for trigger efficiency
for events with a tagged electron (triangles) or positron (squares),
and for all events (circles). 
\label{fig14}}
\end{figure}

The VCS events are selected with the criteria described in Sec.~\ref{evsel}
after applying the $\pi^0$ requirements to the VCS photon. 
These events must satisfy a second trigger line that selects VCS events
with an efficiency close to 100\%, but that is prescaled by a factor of 1000.
The trigger efficiency is 
determined as the fraction of selected events which pass the  
{\tt VirtualCompton} filter. The ratio of
the efficiencies obtained from data and simulation gives the efficiency
correction. We find that the trigger efficiency depends strongly
on electron scattering angle. When $\theta_e$
changes from 0.376 to 0.317 rad, the efficiency falls from 70\% to 30\%
and the efficiency correction increases from 10\% to 30\%. For this reason
the events with $\theta_e < 0.376$ rad were removed from the analysis data
sample.
Since the angular and energy distributions for the VCS and 
$e^+e^-\to e^+e^-\pi^0$
processes are different, we determine the correction as a function of
$\cos{\theta_\gamma}$ and $\cos{\theta_e}$, separately for tagged electrons and
positrons, and convolve this function
with the $\cos{\theta_\pi}$ and $\cos{\theta_e}$ distributions for the
process under study.  The resulting $Q^2$ dependence of the efficiency 
correction is shown in Fig.~\ref{fig14}. The corrections
for events with a tagged electron or positron are also shown. 
The correction for tagged positron events is about $-2\%$ and flat.
For events with a tagged electron, the graph begins at $Q^2 = 7$ GeV$^2$. 
The electron correction changes from $-8\%$ at $Q^2 = 7$ GeV$^2$ to about
$-1.5\%$ at $Q^2 = 20$ GeV$^2$ and higher.

\begin{figure}
\includegraphics[width=.48\textwidth]{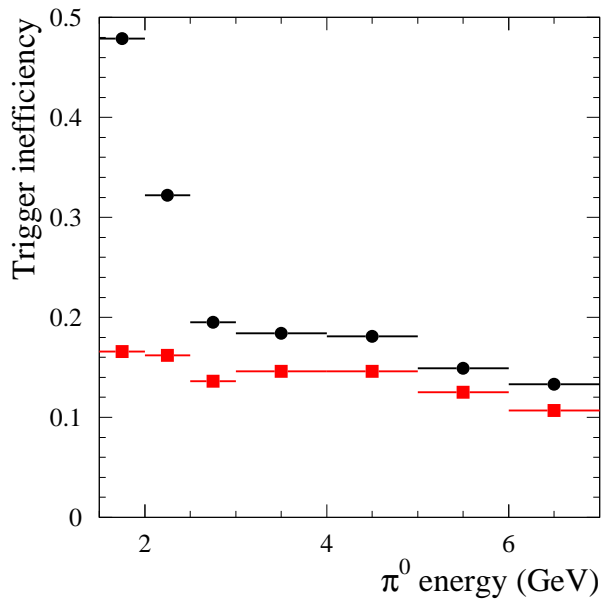}
\caption{
The trigger inefficiency as a function of $\pi^0$ energy determined 
directly from $e^+e^-\to e^+e^-\pi^0$ simulation (circles) and
from simulated VCS events (squares).
\label{fig15}}
\end{figure}
The trigger inefficiency determined directly from 
$e^+e^-\to e^+e^-\pi^0$ simulation is compared to that calculated
using simulated VCS events in Fig.~\ref{fig15}. 
The discrepancy between the inefficiencies is 3-4\% for $\pi^0$ 
energies higher than 3 GeV, but increases to 30\% for $E_\pi < 2$ GeV.
\begin{figure}
\includegraphics[width=.48\textwidth]{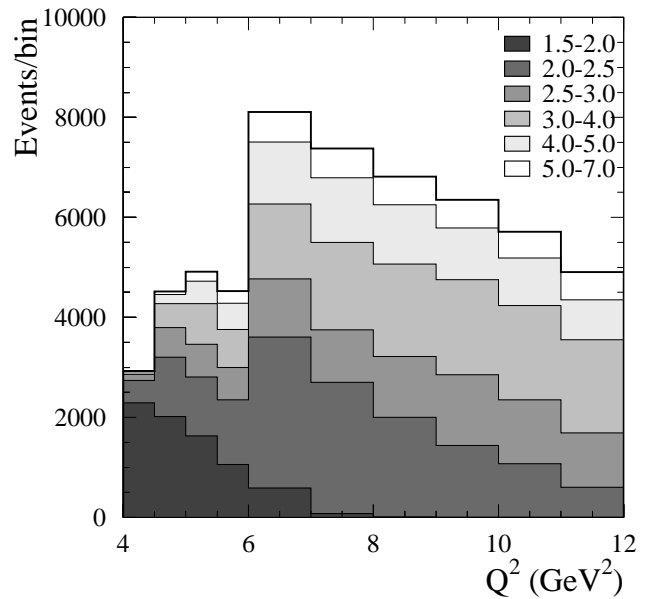}
\caption{The $Q^2$ spectrum for simulated signal events.
The shading represents the contributions from the different $\pi^0$ energy 
ranges (in GeV) indicated by the key.
\label{fig16}}
\end{figure}
Figure~\ref{fig16} shows the $Q^2$ spectra for simulated signal pions of 
different energies.
For each $Q^2$ interval the energy spectra for data and simulation
should be identical. Any difference provides a measure of the quality 
of the trigger efficiency simulation. Using the fitting procedure described
in Sec.~\ref{fitting}, we determine the numbers of signal events in
ten $Q^2$ intervals for six $\pi^0$ energy ranges (53 measurements,
excluding cells with no events). We subtract the $2\pi^0$ background and
normalize the energy spectrum in each $Q^2$ interval so that
its integral is unity.
The same procedure is applied to simulated spectra after introducing
the efficiency corrections for $\pi^0$ loss, EID and trigger inefficiency.
The comparison of the normalized data and simulated spectra gives 
$\chi^2/ndf=42.4/43$ ($ndf=$ number of degrees of freedom).
\begin{figure}
\includegraphics[width=.48\textwidth]{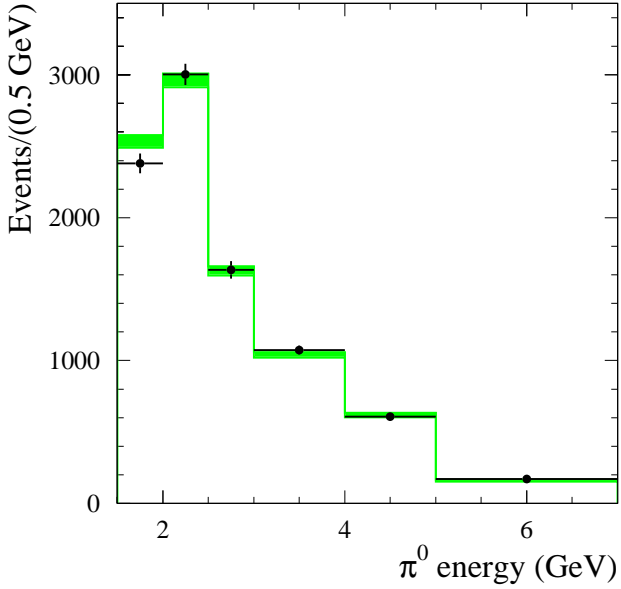}
\caption{The $\pi^0$ energy spectrum for data 
(points with error bars) and signal simulation (histogram). 
The shaded boxes represent the uncertainties associated with the
efficiency correction procedure.
\label{fig17}}
\end{figure}
The $\pi^0$ energy spectrum summed over the $Q^2$ intervals from 4 to 12 GeV$^2$
is shown in Fig.~\ref{fig17}. The simulated spectrum is the sum of
the spectra normalized to the number of data events in each $Q^2$ interval.
The shaded boxes represent the uncertainties in the efficiency corrections. 
The ratio of the data and simulated spectra agrees with unity with
$\chi^2/ndf=5.9/5$. Since the spectra for data and simulation are in 
reasonable agreement, we conclude that the simulation reproduces the 
discrepancy in trigger inefficiency, and so there is no need to introduce an 
extra efficiency correction. However, we introduce an extra systematic 
uncertainty due to trigger inefficiency, which is conservatively estimated to
be 2\%, i.e., half of the difference between the VCS and $e^+e^-\to e^+e^-\pi^0$
trigger inefficiencies for high energy $\pi^0$'s (see Fig.~\ref{fig15}). 

\begin{figure}
\includegraphics[width=.48\textwidth]{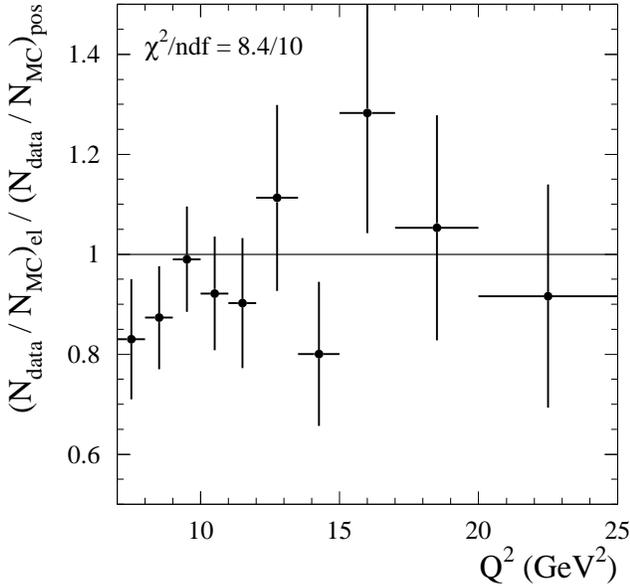}
\caption{The $Q^2$ dependence of the double ratio defined in the text
to compare the cross section for selected signal
events with a tagged electron to that for events with a tagged positron.
\label{fig18}}
\end{figure}
Since the energy and angular distributions, and the trigger efficiency
correction are very different for events with a tagged electron or a tagged
positron, it is interesting to compare the $e^+e^-\to e^+e^-\pi^0$ differential
cross sections measured for electron-tagged and positron-tagged events. 
To do this we subtract the $2\pi^0$ background and apply the efficiency 
correction separately to events with tagged electron and positron. The ratio 
of the cross sections is then calculated as the double ratio
$(N_{\rm{data}}/N_{\rm{MC}})_{\rm{electrons}}/(N_{\rm{data}}/N_{\rm{MC}})_{\rm{positrons}}$
(Fig.~\ref{fig18}) and is found to be in reasonable agreement with unity. 

\subsection{Requirements on $r$ and $\cos{\theta^\ast_{e\pi}}$\label{effcor4}}
To estimate possible systematic uncertainty due to the requirement 
$-0.025<r<0.05$
we study events in the range $0.05<r<0.075$ (see Eq.~(\ref{eqr}) and 
Fig.~\ref{fig3}). 
Assuming that the efficiency corrections are the same for events from
the two $r$ ranges, and subtracting $2\pi^0$ background, we determine the 
double
ratio $(N_{\rm{data}}/N_{\rm{MC}})_{0.05<r<0.075}/(N_{\rm{data}}/N_{\rm{MC}})_{-0.025<r<0.05}$ as
a function of $Q^2$. The ratio is consistent with unity
($\chi^2/ndf=9.3/15$), so we 
conclude that the MC simulation reproduces the shape of the $r$ distribution.

We also study the effect of the $|\cos{\theta^\ast_{e\pi}}|>0.99$ criterion by
changing the value to 0.98 and 0.95. The ratio of the numbers of 
events with $0.98<|\cos{\theta^\ast_{e\pi}}|<0.99$ and 
$|\cos{\theta^\ast_{e\pi}}|>0.99$ is found to be $0.013\pm 0.003$ in 
data and $0.0074\pm 0.0004$ in simulation. The corresponding values
for $0.95<|\cos{\theta^\ast_{e\pi}}|<0.99$ are $0.018\pm 0.002$
and $0.0103\pm 0.0005$. Since the observed data-simulation difference
does not exceed 1\%, we do not introduce an efficiency correction,
but consider this difference (1\%) as a measure of systematic uncertainty
due to the $\cos{\theta^\ast_{e\pi}}$ criterion. 

\begin{figure}
\includegraphics[width=.48\textwidth]{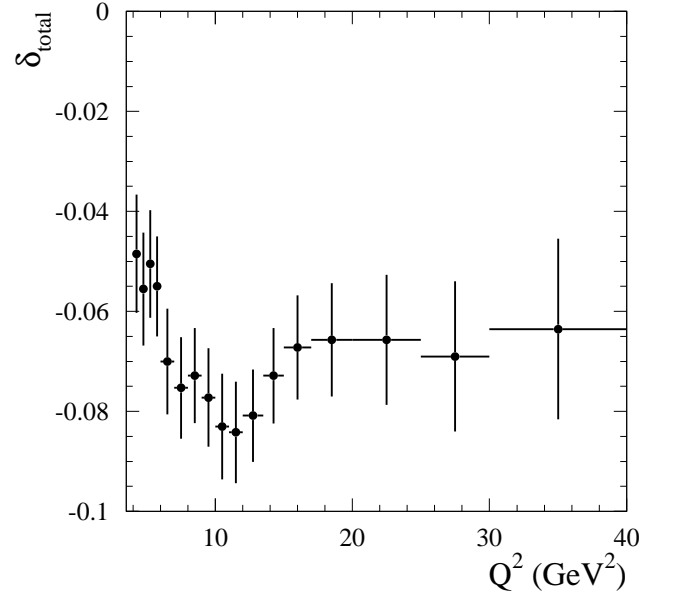}
\caption{The $Q^2$ dependence of the total efficiency correction.
\label{fig19}}
\end{figure}
The total efficiency correction as a function of $Q^2$ is shown in
Fig.~\ref{fig19}, where the error bars are of statistical origin. The
systematic uncertainty, which is independent of $Q^2$, is 2.5\% and
takes into account the uncertainties in the determination of $\pi^0$ loss 
(1\%) and trigger inefficiency (2\%), and the uncertainty due to the
$|\cos{\theta^\ast_{e\pi}}|>0.99$ requirement (1\%). The values of the total
efficiency correction and their statistical errors are listed
in Table~\ref{tab10}.

\section{Cross section and form factor}\label{crosssec}
The Born differential cross section for $e^+e^-\to e^+e^-\pi^0$ is calculated as
\begin{equation}
\frac{\rm{d}\sigma}{\rm{d}Q^2}=
\frac{N_{\rm{cor}}/\Delta Q^2}{\varepsilon RL},
\label{eqcs}
\end{equation}
where $N_{\rm{cor}}$ is the number of signal events corrected for 
data-simulation difference and resolution effects (Table~\ref{tab10}),
$\Delta Q^2$ is the relevant $Q^2$ interval, $L$ is the total 
integrated luminosity (442 fb$^{-1}$), $\varepsilon$ is the detection 
efficiency as a function of $Q^2$, and $R$ is a radiative correction factor 
accounting for distortion of the $Q^2$ spectrum due to the emission of photons
from the initial state particles and for vacuum polarization effects. The 
detection efficiency is obtained from simulation. Its $Q^2$ dependence is shown
in Fig.~\ref{fig4} and listed in Table~\ref{tab10}. 

\begin{figure}
\includegraphics[width=.48\textwidth]{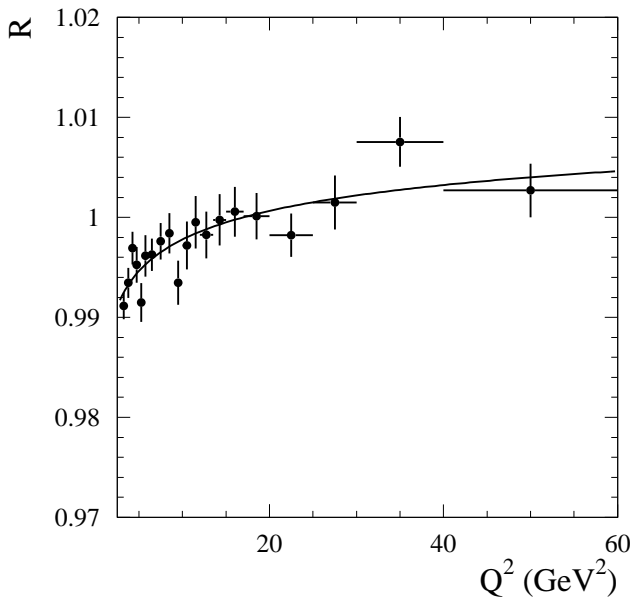}
\caption{The $Q^2$ dependence of the radiative correction factor.
\label{fig20}}
\end{figure}
The radiative correction factor is determined using generator-only simulation.
The $Q^2$ spectrum
is generated using only the Born amplitude for the $e^+e^-\to e^+e^-\pi^0$
process, and then again using a model with radiative corrections included.
The $Q^2$ dependence of the radiative correction factor, evaluated as
the ratio of the second spectrum to the first, is shown in Fig.~\ref{fig20}.
The $Q^2$ dependence is fitted by the function $a/(1+bQ^\gamma)$.
The accuracy of the radiative correction calculation is estimated to 
be 1\%~\cite{RC}.
Note that the value of $R$ depends on the requirement on
the extra photon energy. The $Q^2$ dependence obtained corresponds to
the criterion $r=2E^\ast_\gamma/\sqrt{s}<0.1$ imposed in the simulation.

The corrected mass spectrum ($N_{\rm{cor}}$) is obtained from the 
measured spectrum ($N_{\rm{rec}}$) by dividing by the efficiency
correction factor [see Eq.(\ref{eqcor})] and unfolding the 
effect of $Q^2$ resolution. Using MC simulation,
a migration matrix $A$ is obtained, which represents the probability that
an event with true $Q^2$ in interval $j$ is reconstructed in interval $i$:
\begin{equation}
N_{\rm{rec},i}=
\sum_{j}A_{ij}N_{\rm{cor},j}
\end{equation}
In the case of extra photon emission, $Q^2_{\rm{true}}$ is calculated
as $-(p-p^\prime-k)^2$, where $k$ is the photon four-momentum;
$\varepsilon$ and $R$ in Eq.(\ref{eqcs}) are functions of $Q^2_{\rm{true}}$.
The $Q^2$ resolution varies from about 0.05 GeV$^2$ at $Q^2=5$ GeV$^2$ 
to 0.25 GeV$^2$ at $Q^2=25$ GeV$^2$.
As the chosen $Q^2$ interval size significantly exceeds the resolution
for all $Q^2$, the migration matrix is nearly diagonal, with
 diagonal values $\sim 0.9$, and next-to-diagonal values $\sim 0.05$.
We unfold the $Q^2$ spectrum by applying the inverse of the migration
matrix to the measured spectrum. The procedure changes the shape
of the $Q^2$ distribution insignificantly, but increases the errors
(by about 20\%) and their correlations. The corrected $Q^2$ spectrum 
is listed in Table~\ref{tab10}.

\begin{figure}
\includegraphics[width=.48\textwidth]{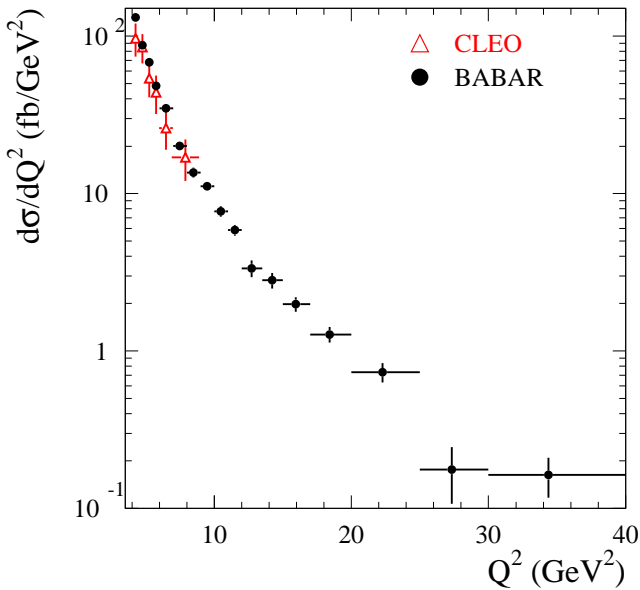}
\caption{The $e^+e^-\to e^+e^-\pi^0$ differential cross section
obtained in this experiment compared to that from the CLEO 
experiment~\cite{CLEO}.
\label{fig21}}
\end{figure}

The values of the differential cross section are listed
in Table~\ref{tab11}. The quoted errors are statistical and systematic.
The latter includes only $Q^2$-dependent errors: the systematic uncertainty in
the number of signal events and the statistical error in the detection 
efficiency determined from MC simulation. The $Q^2$-independent systematic 
error is equal to 3\% and includes the systematic uncertainties in the 
efficiency correction (2.5\%) and in the radiative correction factor (1\%),
and the uncertainty in the integrated luminosity (1\%).
\begin{table}
\caption{The $Q^2$ interval, the weighted average $Q^2$ value for the interval 
$(\overline{Q^2})$, the $e^+e^-\to e^+e^-\pi^0$ cross section
(${\rm{d}\sigma}/{\rm{d}Q^2}(\overline{Q^2})$), and the product of the
$\gamma\gamma^\ast\to \pi^0$ transition form factor $F(\overline{Q^2})$ and
$\overline{Q^2}$. The quoted errors are statistical and systematic for 
the cross section, and combined for the form factor.
In the table we quote only $Q^2$-dependent systematic errors. The 
$Q^2$-independent systematic error is 3\% for the cross section and
2.3\% for the form factor.
\label{tab11}}
\begin{ruledtabular}
\newcolumntype{o}{D{.}{.}{8}}
\begin{tabular}{odod}
\multicolumn{1}{c}{$Q^2$ interval} &
\multicolumn{1}{c}{$\overline{Q^2}$} & 
\multicolumn{1}{c}{${\rm{d}\sigma}/{\rm{d}Q^2}(\overline{Q^2})$} &
\multicolumn{1}{c}{$\overline{Q^2}|F(\overline{Q^2})|$} \\
\multicolumn{1}{c}{(GeV$^2$)} &
\multicolumn{1}{c}{(GeV$^2$)}    &
\multicolumn{1}{c}{(fb/GeV$^2$)}     & 
\multicolumn{1}{c}{(MeV)} \\
\hline 
 4.0-4.5  &  4.24 &  131.4\pm{4.6}\pm{5.0}  &150.4\pm{3.9}\\
 4.5-5.0  &  4.74 &  87.7\pm2.9\pm3.7  &149.1\pm{4.1}\\
 5.0-5.5  &  5.24 &  68.4\pm2.5\pm2.2 &157.4\pm{3.9}\\
 5.5-6.0  &  5.74 &  48.3\pm2.1\pm1.8 &156.0\pm{4.5}\\
 6.0-7.0  &  6.47 &  34.8\pm1.2\pm1.0 &163.5\pm{3.6}\\
 7.0-8.0  &  7.47 & 20.01\pm0.86\pm0.79 &160.6\pm{4.7}\\
 8.0-9.0  &  8.48 & 13.60\pm0.69\pm0.70 &167.3\pm{6.0}\\
 9.0-10.0 &  9.48 & 11.11\pm0.56\pm0.32 &185.3\pm{5.5}\\
10.0-11.0 & 10.48 &  7.73\pm0.48\pm0.38 &186.6\pm{7.6}\\
11.0-12.0 & 11.49 &  5.86\pm0.42\pm0.21 &191.6\pm{7.8}\\
12.0-13.5 & 12.71 &  3.35\pm0.29\pm0.28 &175.\pm{11.}\\
13.5-15.0 & 14.22 &  2.82\pm0.26\pm0.19 &198.\pm{12.}\\
15.0-17.0 & 15.95 &  1.99\pm0.20\pm0.09 &208.\pm{12.}\\
17.0-20.0 & 18.40 &  1.27\pm0.14\pm0.06 &220.\pm{13.}\\
20.0-25.0 & 22.28 &  0.73\pm0.09\pm0.06 &245.\pm{18.}\\
25.0-30.0 & 27.31 &  0.18\pm0.07\pm0.02 &181._{-40.}^{+33.}\\
30.0-40.0 & 34.36 &  0.16\pm0.04\pm0.02 &285._{-45.}^{+39.}\\
\end{tabular}
\end{ruledtabular}
\end{table}
The measured differential cross section at the Born level is shown in 
Fig.~\ref{fig21}, together with CLEO data~\cite{CLEO} for $Q^2>4$ GeV$^2$. 

Because of the strong
nonlinear dependence of the cross section on $Q^2$, the effective value of 
$Q^2$ corresponding to the measured cross section differs from the center of 
the $Q^2$ interval. We parametrize the measured cross section by a smooth
function, reweight the $Q^2$ distribution in simulation to be consistent with 
data,
and calculate the weighted average ($\overline{Q^2}$) for each mass interval.
The values of $\overline{Q^2}$ are listed in Table~\ref{tab11}.

Since the requirement on $\cos{\theta^\ast_{e\pi}}$ limits the momentum 
transfer to the untagged electron, we measure the cross section for
the restricted $q_2^2$ range $|q_2^2| < a_{\rm{max}}$. The value of
$a_{\rm{max}}$
is determined from the  $q^2_2$ dependence of the detection efficiency
($\varepsilon(a_{\rm{max}})=50$\%) and is equal to 0.18 GeV$^2$.

To extract the transition form factor we compare the measured and 
the calculated values of the cross section. The simulation uses 
a constant form factor $F_{\rm{MC}}$. Therefore the measured form factor is
determined from
\begin{equation}
F^2(Q^2)=\frac{(\rm{d}\sigma/\rm{d}Q^2)_{\rm{data}}}
{(\rm{d}\sigma/\rm{d}Q^2)_{\rm{MC}}}F^2_{\rm{MC}}.
\end{equation}
The calculated cross section $(\rm{d}\sigma/\rm{d}Q^2)_{\rm{MC}}$ has
a model-dependent uncertainty due to the unknown dependence on the momentum 
transfer to the untagged electron. We use a $q^2_2$-independent 
form factor, which corresponds to the QCD-inspired model 
$F(q^2_1,q^2_2)\propto 1/(q^2_1+q^2_2)\approx 1/q^2_1$~\cite{Kopp}.
Using the vector dominance model with the form factor $F(q^2_2)\propto
1/(1-q_2^2/m_\rho^2)$, where $m_\rho$ is $\rho$ meson mass,
leads to a decrease of the cross section by 3.5\%.
This difference is considered to be an estimate of model uncertainty due 
to the unknown $q^2_2$
dependence. However, it should be noted that this estimate depends strongly on 
the limit on $q^2_2$. The value of  3.5\% is obtained with $|q^2_2|<0.18$ 
GeV$^2$. For a less stringent $q^2_2$ constraint, for example 
$|q^2_2|<0.6$ GeV$^2$, the difference between the calculated cross sections 
reaches 7.5\%.

\begin{figure}[t]
\includegraphics[width=.48\textwidth]{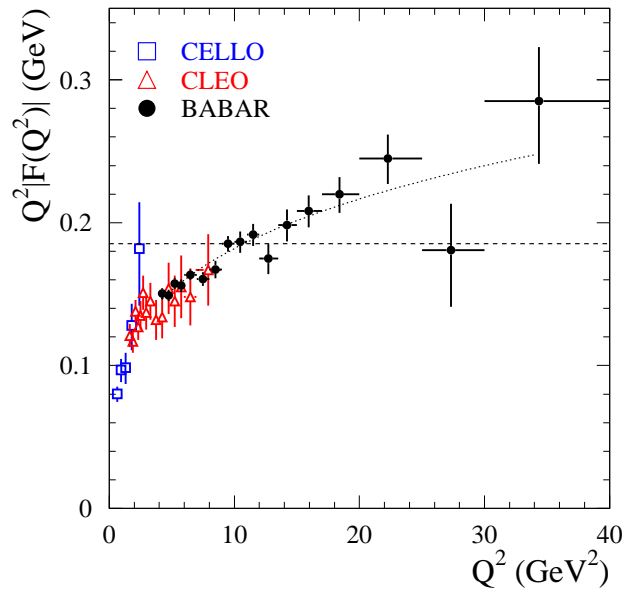}
\caption{The $\gamma\gamma^\ast\to \pi^0$ transition form factor multiplied by
$Q^2$. The dashed line indicates the asymptotic limit for the form factor.
The dotted curve shows the interpolation given by Eq.(\ref{eqinter}).
\label{fig22}}
\end{figure}
The values of the form factor obtained, represented in the form 
$\overline{Q^2}|F(\overline{Q^2})|$, are listed in Table~\ref{tab11}
and shown in Fig.~\ref{fig22}.
For the form factor we quote the combined error, for which
the statistical and $Q^2$-dependent systematic uncertainties are
added in quadrature. The $Q^2$-independent systematic error is 
2.3\%, and includes the uncertainty on the measured differential
cross section, and the model-dependent uncertainty due to the unknown $q_2^2$ 
dependence.

\section{Conclusions}
We have studied the $e^+e^- \to e^+e^-\pi^0$ reaction in the
single tag mode and measured the differential cross section
$(\rm{d}\sigma/\rm{d}Q^2)$ and the $\gamma\gamma^\ast\to \pi^0$ transition 
form factor $F(Q^2)$ for the momentum transfer range from 4 to
40 GeV$^2$. For the latter, the comparison of our results with previous
measurements~\cite{CELLO,CLEO} is shown in Fig.~\ref{fig22}. 
In the $Q^2$ range from 4 to 9 GeV$^2$ our results are in 
reasonable agreement with the measurements by the CLEO collaboration~\cite{CLEO}, but have
significantly better precision. We also significantly extend the
$Q^2$ region over which the form factor is measured.
\begin{figure}
\includegraphics[width=.48\textwidth]{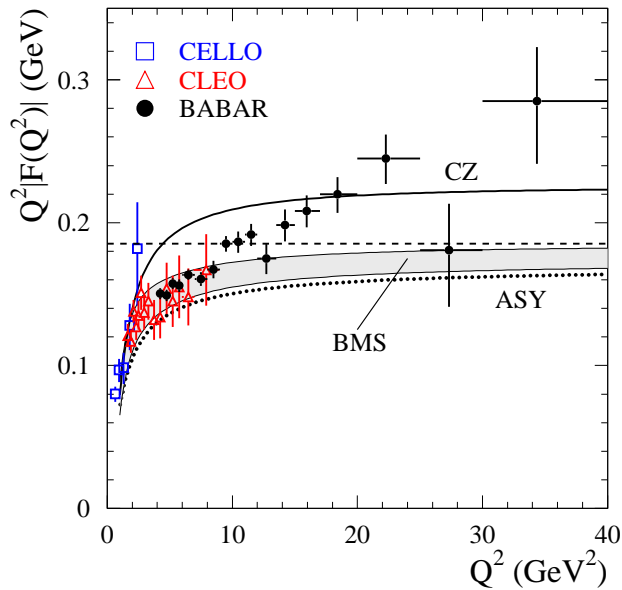}
\caption{The $\gamma\gamma^\ast\to \pi^0$ transition form factor multiplied by
$Q^2$. The dashed line indicates the asymptotic limit for the form factor.
The solid and dotted lines show the predictions for the form factor~\cite{th6}
for the CZ~\cite{CZ} and asymptotic (ASY)~\cite{ASY} models for the pion
distribution amplitude, respectively. The shaded band represents 
the prediction for the BMS~\cite{BMS} pion DA model.
\label{fig23}}
\end{figure}

To effectively describe the $Q^2$ dependence of
the form factor in the range 4--40 GeV$^2$, we fit
the function
\begin{equation} 
Q^2|F(Q^2)|=A\left( \frac{Q^2}{10\mbox{ GeV}^2}\right)^\beta 
\label{eqinter}
\end{equation}
to our data. The values obtained for the parameters are $A=0.182\pm0.002$ GeV, 
and $\beta=0.25\pm0.02$. The fit result is shown in Fig.~\ref{fig22}
by the dotted curve. The effective $Q^2$ dependence of the form factor
($\sim 1/Q^{3/2}$) differs significantly from the leading order pQCD prediction
($\sim 1/Q^{2}$) (see Eq.(\ref{LO})), demonstrating the importance of
higher-order pQCD and power corrections in the $Q^2$ region under study.

The horizontal dashed line in Fig.~\ref{fig22} indicates the asymptotic
limit $Q^2F(Q^2)=\sqrt{2}f_\pi\approx 0.185$ GeV for $Q^2 \to \infty$, 
predicted by pQCD~\cite{LB}. The measured
form factor exceeds the limit for $Q^2 > 10$ GeV$^2$. This contradicts
most models for the pion distribution amplitude 
(see, e.g., Ref.~\cite{Stefanis} and 
references therein), which give form factors approaching the asymptotic limit 
from below.

The comparison of the form factor data to the predictions of some theoretical
models is shown in Fig.~\ref{fig23}. The calculation of~\cite{th6} was performed 
by A.~P.~Bakulev, S.~V.~Mikhailov and 
N.~G.~Stefanis using the light-cone sum rule method~\cite{th9,th2} 
at next-to-leading order (NLO) pQCD; the power correction due to the
twist-4 contribution~\cite{th9} was also taken into account. Their results are shown 
for the Chernyak-Zhitnitsky DA (CZ)~\cite{CZ}, the asymptotic DA (ASY)~\cite{ASY}, 
and the DA derived from QCD sum rules with non-local 
condensates (BMS)~\cite{BMS}. 

For all three DAs the $Q^2$ dependence is almost flat for $Q^2 \gtrsim 10$
GeV$^2$, whereas the data show significant growth of the form factor
between 8 and 20 GeV$^2$. This indicates that the NLO pQCD approximation with
twist-4 power correction, which has been widely used for the description of 
the form-factor measurements by the CLEO collaboration~\cite{CLEO},
is inadequate for $Q^2$ less than $\sim15$ GeV$^2$. 
In the $Q^2$ range from 20 to 40 GeV$^2$, where uncertainties due to higher 
order pQCD and power corrections are expected to be smaller,
our data lie above the asymptotic limit and are consistent
with the CZ model.

\begin{acknowledgments}
We thank V.~L.~Chernyak for useful discussions, and
A.~P.~Bakulev, S.~V.~Mikhailov, and N.~G.~Stefanis for
providing us the NLO pQCD calculation of the transition 
form factor. We are grateful for the                                                 
extraordinary contributions of our \pep2\ colleagues in                 
achieving the excellent luminosity and machine conditions               
that have made this work possible.                                      
The success of this project also relies critically on the               
expertise and dedication of the computing organizations that            
support \babar.                                                         
The collaborating institutions wish to thank                            
SLAC for its support and the kind hospitality extended to them.         
This work is supported by the                                           
US Department of Energy                                                 
and National Science Foundation, the                                    
Natural Sciences and Engineering Research Council (Canada),             
the Commissariat \`a l'Energie Atomique and                             
Institut National de Physique Nucl\'eaire et de Physique des Particules 
(France), the                                                           
Bundesministerium f\"ur Bildung und Forschung and                       
Deutsche Forschungsgemeinschaft                                         
(Germany), the                                                          
Istituto Nazionale di Fisica Nucleare (Italy),                          
the Foundation for Fundamental Research on Matter (The Netherlands),    
the Research Council of Norway, the                                     
Ministry of Science and Technology of the Russian Federation,           
Ministerio de Educaci\'on y Ciencia (Spain), and the                    
Science and Technology Facilities Council (United Kingdom).             
Individuals have received support from                                  
the Marie-Curie IEF program (European Union) and                        
the A. P. Sloan Foundation.                   
\end{acknowledgments} 


\end{document}